\documentclass{IEEEtran}
\usepackage{cite}
\usepackage{graphicx}
\usepackage{mathrsfs}
\usepackage{stfloats}
\usepackage{dsfont}
\usepackage{setspace}
\usepackage{array}
\usepackage{bbm}
\usepackage{hyperref}
\usepackage{here}
\usepackage{amsmath,amsthm,amsfonts,amssymb}
\usepackage{color}
\usepackage{accents}

\usepackage{epstopdf}
\usepackage[centerlast,small]{caption}
\usepackage{subcaption}
\usepackage{bbm}
\usepackage{easytable}
\usepackage{booktabs}

\usepackage{fancyvrb} 
\VerbatimFootnotes    

\usepackage{algorithm,algorithmic}

\begin{document}

\title{Bi-objective Optimization of Information Rate and Harvested Power in RIS-aided SWIPT Systems}
\author{Abdelhamed Mohamed, A. Zappone,~\IEEEmembership{Senior Member,~IEEE}  and Marco~Di~Renzo,~\IEEEmembership{Fellow,~IEEE}
\thanks{Manuscript received April 22, 2022. A. Mohamed and M. Di Renzo are with Universit\'e Paris-Saclay, CNRS and CentraleSup\'elec, Laboratoire des Signaux et Syst\`emes,  91192 Gif-sur-Yvette, France. A. Zappone is with the University of Cassino and Southern Lazio, 03043 Cassino,  Italy. (e-mail: marco.direnzo@centralesupelec.fr). }}

%
%
%
%
\maketitle
\begin{abstract}
The problem of simultaneously optimizing the information rate and the harvested power in a reconfigurable intelligent surface (RIS)-aided multiple-input single-output downlink wireless network with simultaneous wireless information and power transfer (SWIPT) is addressed. The beamforming vectors, RIS reflection coefficients, and power split ratios are jointly optimized subject to maximum power constraints, minimum harvested power constraints, and realistic constraints on the RIS reflection coefficients. A practical algorithm is developed through an interplay of alternating optimization, sequential optimization, and pricing-based methods. Numerical results show that the deployment of RISs can significantly improve the information rate and the amount of harvested power.
\end{abstract}

\section{Introduction} \label{Introduction}
Reconfigurable Intelligent Surfaces (RISs) have emerged as a promising technology for sustainable 6G networks \cite{RIS_GE_JSAC,Wu2019f}. Thanks to their ability of reflecting and refracting electromagnetic signals in a reconfigurable fashion and with limited energy requirements, RISs can drastically reduce the energy consumptions in wireless networks\cite{huang2019holographic}. In this context, RISs have been also studied in conjunction with the use of simultaneous wireless information and power transfer (SWIPT), which is another key technology to improve the energy-sustainability of future wireless networks.

Several studies show that the use of an RIS can improve both information and power transfer. In~\cite{Wu2019k}, the problem of transmit power minimization subject to quality of service (QoS) constraints and minimum energy harvesting requirements is addressed. The optimization problem is tackled by means of a penalty-based algorithm coupled with the alternating optimization technique. In \cite{Z.LiNOMASWIPT}, the problem of transmit power minimization for an RIS-assisted SWIPT non-orthogonal multiple-access (NOMA)  network is investigated. A two-stage optimization algorithm is proposed to jointly optimize the transmit beamforming vector, the power-split  ratio,  and the RIS  phase shifts under QoS constraints. Semidefinite relaxation coupled with alternating and sequential optimization methods are employed. In~\cite{Pan2019b}, the problem of maximizing the weighted sum rate maximization is investigated in a SWIPT-based multi-user multiple-input multiple-output (MIMO) downlink system, subject to minimum harvested energy constraints. Alternating optimization is used in conjunction with sequential optimization and pricing methods.
In \cite{xu2021optimal}, the authors study the problem of resource allocation in RIS-aided SWIPT-based systems, in which a large RIS is divided into several tiles to be separately designed with the objective of reducing the computational complexity of the design. Both a globally optimal algorithm and a practical approach are developed by means of branch-and-bound and sequential methods.
In \cite{khalili2021multiobjective}, the trade-off between sum-rate maximization and the total harvested energy is investigated. The $\epsilon-$method coupled with alternating optimization is used to tackle the resulting multi-objective problem.
In \cite{Chen_He_A_Joint_Power_Splitting}, the data rate maximization problem in an RIS-aided system in which multiple receivers perform both information decoding and wireless power reception is analyzed. The problem is tackled by alternating optimization, sequential optimization, and sub-gradient searches.

This work considers a network in which a multiple-antenna base station serves single-antenna users with the aid of an RIS. Each receiver jointly performs information decoding and wireless power reception by means of power splitting. Unlike previous works, the following contributions are made.

\textbf{1)} We consider the bi-objective problem of simultaneously maximizing the information rate and the harvested power. The problem is formulated subject to minimum rate constraints on the downlink channel, minimum harvested power, RIS phase constraints, and maximum transmit power constraint. The resulting NP-hard problem is tackled by an interplay of alternating maximization, sequential optimization, and penalty-based methods. Since the receivers perform both information decoding and wireless power reception, the optimization of the power split ratio is needed, which a novel feature compared to related works on RIS-aided, SWIPT-based systems.

\textbf{2)} The work considers the realistic case in which the phases and moduli of the RIS reflection coefficients are not independent with one another, but are rather coupled by a deterministic function. This further complicates the solution of the resource allocation problem.

\textbf{3)} Numerical results confirm the effectiveness of the proposed algorithm compared to traditional approaches. It is found, in particular, that increasing the number of RIS elements significantly enhances the achieved data rate and the amount of harvested power.

Among previous works, \cite{khalili2021multiobjective} and  \cite{Chen_He_A_Joint_Power_Splitting} are the most closely-related to our work. However, \cite{khalili2021multiobjective} investigates the rate and  harvested energy trade-off for separated information and power receivers, i.e., each receiver performs either information decoding or wireless power harvesting. Also, independent phases and moduli are assumed for the RIS reflection coefficients. In addition, \cite{Chen_He_A_Joint_Power_Splitting} considers integrated information and power receivers, but it focuses on maximizing only the information rate, without considering power harvesting and the intertwinement between the phase and the amplitude of the RIS reflection coefficients.

\begin{figure}[!t]
\centering
\includegraphics[width=0.5\textwidth,height=0.5\textheight,keepaspectratio]{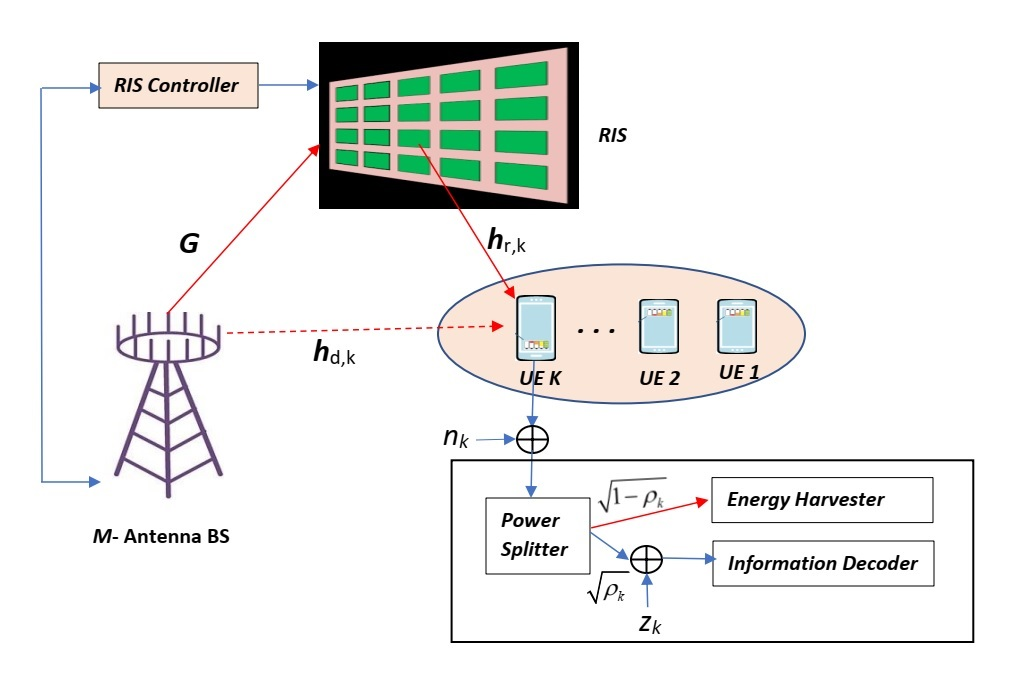}
	\caption{Illustration of the MISO RIS-assisted SWIPT system model.} \label{System_model}
\end{figure}

\section{System Model and Problem Formulation}\label{sec:System Model and Problem formulation}
We consider an RIS-based multi-user multiple-input single-output (MISO) downlink system in which a base station (BS) equipped with $M$ antennas serves $K$ single-antenna user equipments (UEs) employing SWIPT. The $k$-th UE employs a fraction $\rho_{k}$ of the received power for information decoding (ID), while the rest is used for power harvesting (PH). The channels from the BS to the RIS, from BS to \textit{k}-th user, and from the RIS to the \textit{k}-th user are denoted by $\mathbf{G}\in {{\mathbb{C}}^{N\times M}}$, ${{\mathbf{h}}_{d,k}}\in {{\mathbb{C}}^{1\times M}}$, and ${\mathbf{h}_{r,k}}\in {{\mathbb{C}}^{1\times N}}$, respectively, and are assumed to follow the block Rician fading model. The reflection coefficient vector of the RIS is defined by $\mathbf{v} = [{v_1},...,{v_n}]\in {{\mathbb{C}}^{N\times 1}}$, where ${v_n} = {f_n}({\theta _n}){e^{j{\theta _n}}}$ is the reflection coefficient of the \textit{n}-th reflecting element of the RIS, with $ - \pi  \le {\theta _n} \le \pi $ and, e.g., ${f_n}({\theta _n}) = {f_{\min }} + (1 - {f_{\min }}){\left( {\frac{{\sin ({\theta _n} - \phi ) + 1}}{2}} \right)^\alpha }$ is a function that relates the phase of the reflection coefficient to its modulus, where ${f_{\min }} \geqslant 0$, $\alpha  \geqslant 0, \phi  \geqslant 0$ are circuit implementation constants \cite{Abeywickrama2020a}. In general, the proposed approach can be applied to any continuous and differentiable function ${f_n}({\theta _n})$. Given the above notation, and defining $\mathbf{H}_{r,k}=\text{diag}(\mathbf{h_{r,k}})\mathbf{G}$, the achievable sum-rate is:
\begin{align}
&{R^{ID}}\!=\!\sum_{k=1}^{K}\log \!\left(\!1 \!+\! \frac{{{{\left| {\left( {{\mathbf{h}_{d,k}} + {\mathbf{v} ^H}{\mathbf{H}_{r,k}}} \right){\mathbf{w}_k}} \right|}^2}}}{{\sum\limits_{i = 1,i \ne k}^{{K}}\!\! {{{\left| {\left( {{\mathbf{h}_{d,k}} \!+\! {\mathbf{v} ^H}{\mathbf{H}_{r,k}}} \right){\mathbf{w}_i}} \right|}^2}\!+\!{\sigma_{k}^2}\!+\!\frac{\delta_{k} ^2}{\rho_{k}} }}}\right)
\end{align}
wherein ${{\mathbf{w}_k}}\in {{\mathbb{C}}^{M\times 1}}$ is the transmit beamforming vector, while $\sigma^2_{k}$ and $\delta^2_{k}$ model the power of the thermal noise and of the noise due to the conversion of the RF signal to the baseband. Furthermore, considering a linear harvesting model, the power harvested by the $k$-th UE is $P_{H,k}=\eta _k(1 - {\rho _k})\sum\nolimits_{i = 1}^K {\left| {(\mathbf{h}_{d,k} + \mathbf{v} ^H\mathbf{H}_{r,k}){\mathbf{w}_i}} \right|}^2$, where $\eta_k \in [0,1]$ is the efficiency of the power harvesting circuit. Similar to \cite{Tang2020}, we consider the associated rate function:
\begin{equation}
  \begin{aligned}
 &R^{PH}\!=\!\sum\limits_{k = 1}^K\! \log \!\left(\! 1 + \frac{\xi_{k}\eta _k(1 - {\rho _k})}{\sigma _k^2}\sum\limits_{i = 1}^K {\left| {(\mathbf{h}_{d,k} + \mathbf{v} ^H\mathbf{H}_{r,k}){\mathbf{w}_i}} \right|}^2 \!\right)\label{equivalent EH_to_rate}
  \end{aligned}
 \end{equation}
where $\xi_k \in [0,1]$ is the efficiency of the conversion from baseband power to RF power.

The goal of this work is to maximize a weighted sum of $R^{ID}$ and  $R^{PH}$, namely:
  \begin{equation}
   \begin{aligned}
R^{Eq}_{sum}{(\rm{\rho},\mathbf{w},\mathbf{v},\left\lbrace \theta_n\right\rbrace)} &=R^{ID}+ \bar{\lambda}R^{PH}\label{Equivalent_ESR}
 \end{aligned}
 \end{equation}
where $\bar{\lambda}$ is a parameter to be tuned by the network operator according to the priorities granted to ID and PH \cite{Tang2020}.

Defining $\mathbf{h}_{k}=\mathbf{h}_{d,k} + \mathbf{v} ^H\mathbf{H}_{r,k}$, the problem to be tackled in the rest of this work is formulated as:
\begin{align}
\mathcal {P_{A}}:&\quad \mathop {\mathtt{max} }\limits_{\rm{\rho},\mathbf{w},\mathbf{v},\left\lbrace \theta_n\right\rbrace} {\rm{ }} R^{Eq}_{sum}{(\rm{\rho},\mathbf{w},\mathbf{v},\left\lbrace \theta_n\right\rbrace)}\\
\mathtt{s.t.} &\text{C1:}\;\frac{{{{\left| { \mathbf{h}_{k}{\mathbf{w}_k}} \right|}^2}}}{{\sum\limits_{i \ne k}^{{K}} {{{\left| {\mathbf{h}_{k}{\mathbf{w}_i}} \right|}^2}\!+\!{\sigma_{k}^2}\!+\!\frac{\delta_{k} ^2 }{{\rho_{k}}}}}}\ge \gamma_{min}, \;k=1,\ldots,K \\
&\text{C2:}~{\eta _k}(1 \!-\! {\rho _k})\sum\limits_{i = 1}^K {{{\left| {{\mathbf{h}_k}{\mathbf{w}_i}} \right|}^2}}  \ge P_{min},\;k=1,\ldots,K\\
&\text{C3:}~\sum\limits_{k = 1}^{{K}} {{{\left\| {{\mathbf{w}_k}} \right\|}^2} \le {P_T},}\;,\;~ 0 \le \rho_{k}\le 1 \\
&\text{C4:}~{v_n} = {f_n}({\theta _n}){e^{j{\theta _n}}},\quad{\rm{  }}n = 1,\dots,N\\
&\text{C5:}~ - \pi  \le {\theta _n} \le \pi,\quad{\rm{  }}~n = 1,\dots,N
\end{align}

It can be seen that $\mathcal{P_{A}}$ is a non-convex problem due to the non-concavity of both the objective function and the constraints C1, C2, C4. Thus, traditional methods do not apply.

\section{Proposed Approach}
To tackle $\mathcal{P_{A}}$, we first reformulate the sum of logarithms into a more tractable form by applying the method from \cite{Shen2018} to each sum in our objective function. This yields:
 \begin{equation}
 \begin{aligned}
 \overline {{\mathcal{P}_\mathcal{A}}} :&\mathop {{\mathtt{max}}}\limits_{\scriptstyle{\rm\alpha _I},{\rm\beta _I},{\rm\alpha _E},{\rm\beta _E},\scriptstyle{\rm{\rho},\mathbf{w},\mathbf{v},\left\lbrace \theta_n\right\rbrace}} {f_\mathcal{A}}({\rm\alpha _I},{\rm\beta _I},{\rm\alpha _E},{\rm\beta _E},{\rm{\rho},\mathbf{w},\mathbf{v},\left\lbrace \theta_n\right\rbrace})\\~~~&\text{s.t}~\text{(C1)},\text{(C2)},\text{(C3)},\text{(C4)},\text{(C5)}\label{equivalentreformulatedoptimizationproblem}
 \end{aligned}
 \end{equation}
wherein ${f_\mathcal{A}}$ is shown in \eqref{equivalentreformulatedoptimizationfunction} at the top of the next page, with ${{\bar \eta }_k} = {\xi _k}{\eta _k}$ and $\Re$ being the real part operator.
\begin{figure*}
  \begin{equation}
  \begin{aligned}
 & {f_\mathcal{A}}({\rm\alpha _I},{\rm\beta_I},{\rm\alpha_E},{\rm\beta _E},{\rm{\rho},\mathbf{w},\mathbf{v},\left\lbrace \theta_n\right\rbrace} ) = \sum\limits_{k = 1}^K {\log (1 + {\alpha_{I,k}}) + \bar \lambda \log (1 + {\alpha _{E,k}}) - \left( {{\alpha _{I,k}} + \bar \lambda {\alpha _{E,k}}} \right)}\\&
  + \sum\limits_{k = 1}^K {\left( {2\sqrt {{\rho _k}(1 + {\alpha_{I,k}})} \Re (\beta _{I,k}^*{\mathbf{h}_k}{\mathbf{w}_k})} \right)}  - \sum\limits_{k = 1}^K {\left( {{{\left| {{\beta _{I,k}}} \right|}^2}\left( {\sum\limits_{i = 1}^K {{\rho _k}{{\left| {{\mathbf{h}_k}{\mathbf{w}_i}} \right|}^2} + {\rho _k}{\sigma _k}^2}  + \delta _k^2} \right)} \right)} \\
  &+ \sum\limits_{k = 1}^K {\left( {2\bar \lambda \sqrt {{{\bar \eta }_k}(1 - {\rho _k})(1 + {\alpha _{E,k}})} \Re (\beta _{E,k}^*\sum\limits_{i = 1}^K {{\mathbf{h}_k}{\mathbf{w}_i}} )} \right)}- \sum\limits_{k = 1}^K {\left( {\bar \lambda {{\left| {{\beta _{E,k}}} \right|}^2}\left( {{\xi _k}{\eta _k}(1 - {\rho _k})\sum\limits_{i = 1}^K {{{\left| {{{\mathbf{h}}_k}{{\mathbf{w}}_i}} \right|}^2}}  + {\sigma _k}^2} \right)} \right)}
   \label{equivalentreformulatedoptimizationfunction}
  \end{aligned}
  \end{equation}
\end{figure*}
In order to tackle \eqref{equivalentreformulatedoptimizationproblem}, the first step is to embed (C4) into the objective, resorting to a penalty-based approach, which yields:
 \begin{equation}
  \begin{aligned}
 	\overline {{\mathcal{P}_\mathcal{B}}} :&\mathop {{\mathtt{max}}}\limits_{\scriptstyle{\rm\alpha _I},{\rm\beta _I},{\rm\alpha _E},{\rm\beta _E},\scriptstyle{\rm{\rho},\mathbf{w},\mathbf{v},\left\lbrace \theta_n\right\rbrace}} {f_\mathcal{A}} - \Gamma \sum\limits_{n = 1}^N {{{\left| {{v_n} - {f_n}({\theta _n}){e^{j{\theta _n}}}} \right|}^2}}\\~~~&\text{s.t}~\text{(C1)},\text{(C2)},\text{(C3)},\text{(C5)}\label{penalizedoptimizationproblem}
 \end{aligned}
\end{equation}
wherein $ \Gamma$ represents the penalty coefficient used for penalizing the violation of the equality constraint (C4). If  $\Gamma  \to \infty $, the solution of the original problem is obtained. Problem \eqref{penalizedoptimizationproblem} will be tackled by alternating optimization, as explained next.

\subsection{Optimization of ${\alpha_{I,k}},{\alpha _{E,k}}$, $\beta_{Ik},\beta_{Ek},\rho_{k}$}
The optimal ${\alpha_{I,k}},{\alpha _{E,k}},\beta_{Ik},\beta_{Ek}$ are found by simply setting the  gradient of the objective to zero, which yields:
\begin{align}
&{\alpha_{I,k}}\!=\!\frac{{{r^2} \!+\! r\sqrt {{r^2} \!+\! 4} }}{2},{\beta _{I,k}}\! =\! \frac{{\sqrt {{{ \rho }_k}(1\! +\! {{\alpha }_{Ik}})} ({{{\mathbf{h}}}_k}{{ {\mathbf{w}}}_k})}}{{\!\sum\limits_{i = 1}^K {{{ \rho }_k}{{\left| {{{{\mathbf{h}}}_k}{{ {\mathbf{w}}}_i}} \right|}^2}\!\! +\! {{ \rho }_k}{\sigma _k}^2}\! + \!\delta _k^2}}\label{alpha_beta_I}
\end{align}
\begin{align}
& {\alpha _{E,k}}\!=\!\frac{{{{\tilde r}^2} \!+\! \tilde r\sqrt {{{\tilde r}^2} \!+\! 4} }}{2}, {\beta _{E,k}}\! =\! \frac{{\sqrt {{{\bar \eta }_k}(1\! -\! {{ \rho }_k})(1 \!+\! {{ \alpha }_{Ek}})}\! \sum\limits_{i = 1}^K {{{ {\mathbf{h}}}_k}{{ {\mathbf{w}}}_i}} }}{{ {{{\bar \eta }_k}(1\! -\! {{ \rho }_k})\sum\limits_{i = 1}^K {{{\left| {{{ {\mathbf{h}}}_k}{{ {\mathbf{w}}}_i}} \right|}^2}} \! +\! {\sigma _k}^2}}}\label{alpha_beta_E}
\end{align}
with $\tilde r\!\!=\!\!\sqrt {\bar \eta (1 \!-\! {\rho _k})} \Re \!\left\{\! {\beta _{E,k}^ *\! \sum\limits_{i = 1} {{{{{\mathbf{h}}}_k}{{ {\mathbf{w}}}_i}}} } \!\right\}$, $r\!\! =\!\! \sqrt {{\rho _k}} \Re \!\left\{\! {\beta _{I,k}^ * {{{{{\mathbf{h}}}_k}{{ {\mathbf{w}}}_k}}}} \!\right\}$.

The optimization with respect to the coefficients $\{\rho_{k}\}$ is also straightforward. With respect to $\{\rho_{k}\}$, in fact, the objective function is strictly concave and the constraints (\text{C2}), (\text{C3}) are affine. Moreover, (\text{C1})  can be rewritten in a linear form as follows, for any $k=1,\ldots,K$:
  \begin{equation}
  {\rho _k}{\left| {{{ {\mathbf{h}}}_k}{{ {\mathbf{w}}}_k}}\right|^2} - {\gamma _{min}}\left( {{\rho _k}\sum\limits_{i = 1,i \ne k}^K {{{\left| {{{ {\mathbf{h}}}_k}{{ {\mathbf{w}}}_i}} \right|}^2} + {\rho _k}\sigma _k^2 + \delta _k^2} } \right) \ge 0 \label{ps_c1}
  \end{equation}
Thus, with respect to $\{\rho_{k}\}$, the problem is convex and can be solved by standard convex optimization algorithms \cite{993483}.

\subsection{Optimization of $\mathbf{w}_{k}$}
When all the other variables are fixed, the objective function is a concave function of the transmit beamforming vectors $\mathbf{w}_{k}$. However, constraints (C1) and (C2) are still not convex.  To deal with them, we resort to the successive convex approximation (SCA) method \cite{Pan2019b},\cite{8579566}. Specifically, the convex term ${\left| {{{ {\mathbf{h}}}_k}{{\mathbf{w}}_k}} \right|}^2$ is upper-bounded by its first-order Taylor expansion as follows:
 \begin{equation}\label{Eq:Approx1}
{\mathbf{w}}_k^H{\mathbf{h}}_k^H{{\mathbf{h}}_k}{{\mathbf{w}}_k}\! \geq \!2\Re \!\left(\! {{\mathbf{w}}_k^{(t)H}\!{\mathbf{h}_k^H}{{\mathbf{h}}_k}{{\mathbf{w}}_k}} \!\right) \!\!-\!\! \left(\! {{\mathbf{w}}_k^{(t)H}\!{\mathbf{h}}_k^H{{\mathbf{h}}_k}{\mathbf{w}}_k^{(t)}} \!\right)
 \end{equation}
whereing ${{\mathbf{w}}_k^{(t)},\forall k}$ is the solution from the previous iteration. Thus, exploiting \eqref{Eq:Approx1} and elaborating, (C1) can be recast as:
 \begin{equation}
\begin{aligned}
&\gamma_{\min}\!\left(\!\sum_{i\neq k}|\mathbf{h}_{k}\mathbf{w}_{i}|^{2}\!+\!\sigma_{k}^{2}\!+\!\frac{\delta_{k}^{2}}{\rho_{k}}\right)\\
&\!+\! {{\mathbf{w}}_k^{(t)H}{\mathbf{h}}_k^H{{\mathbf{h}}_k}{\mathbf{w}}_k^{(t)}}\!\!-\!2\Re \left( {{\mathbf{w}}_k^{(t)H}{\mathbf{h}_k^H}{{\mathbf{h}}_k}{{\mathbf{w}}_k}} \right) \!\leq\! 0
 \end{aligned}\label{W_C1}
 \end{equation}
 
Similarly, (C2) can be reformulated as follows:
 \begin{align}
\begin{aligned}
\eta_{k}(1\!-\!\rho_{k})\!&\sum_{k=1}^{K}2\Re \left( {{\mathbf{w}}_i^{(t)H}{\mathbf{h}_k^H}{{\mathbf{h}}_k}{{\mathbf{w}}_i}} \right)
\!-\! {{\mathbf{w}}_i^{(t)H}{\mathbf{h}}_k^H{{\mathbf{h}}_k}{\mathbf{w}}_i^{(t)}} \\ & \hspace{1.5cm} \!\geq \!P_{min}
 \end{aligned}  \label{W_C2}
 \end{align}

By replacing the constraints (C1) and (C2) with \eqref{W_C1} and \eqref{W_C2}, we obtain the convex surrogate problem to be solved in each iteration of the SCA method for optimizing $\mathbf{w}_{k}$.

\subsection{Optimization of $\mathbf{v}$}
The approach is similar to that used for the optimization of $\mathbf{w}_{k}$. In fact, the objective is concave in $\mathbf{v}$, while constraints (C1) and (C2) can be handled by the SCA method. Specifically, (C1) can be replaced by the convex constraint:
\begin{align}\label{V_C1}
&\gamma_{\min }\left(\sum\limits_{i \neq k}^K |(\mathbf{h}_{d,k}+\mathbf{v}^{H}\mathbf{H}_{r,k})\mathbf{w}_{i}|^{2}+\sigma_{k}^{2}+\frac{\delta_{k}^{2}}{\rho_{k}}\right)\\
&+(\mathbf{h}_{d,k}+\mathbf{v}^{(t)H}\mathbf{H}_{r,k})\mathbf{w}_{k}\mathbf{w}_{k}^{H}(\mathbf{h}_{d,k}+\mathbf{v}^{(t)H}\mathbf{H}_{r,k})^{H}\notag\\
&-2\Re\{(\mathbf{h}_{d,k}+\mathbf{v}^{(t)H}\mathbf{H}_{r,k})\mathbf{w}_{k}\mathbf{w}_{k}^{H}(\mathbf{h}_{d,k}+\mathbf{v}^{H}\mathbf{H}_{r,k})^{H}\}\leq 0\notag\;
\end{align}
and (C2) by the convex constraint:
\begin{align}\label{V_C2}
&\eta_{k}(1\!-\!\rho_{k})\!\sum_{i=1}^{K}(\mathbf{h}_{d,k}\!+\!\mathbf{v}^{(t)H}\mathbf{H}_{r,k})\mathbf{w}_{i}\mathbf{w}_{i}^{H}(\mathbf{h}_{d,k}\!+\!\mathbf{v}^{(t)H}\mathbf{H}_{r,k})^{H}\!\!-\notag\\
&2\Re\{\!(\mathbf{h}_{d,k}\!\!+\!\mathbf{v}^{(t)H}\!\mathbf{H}_{r,k})\mathbf{w}_{i}\mathbf{w}_{i}^{H}(\mathbf{h}_{d,k}\!\!+\!\mathbf{v}^{H}\!\mathbf{H}_{r,k})^{H}\!\}\!\geq\! P_{min}
\end{align}

By replacing the constraints (C1) and (C2) with \eqref{V_C1} and \eqref{V_C2}, we obtain the convex surrogate problem to be solved in each iteration of the SCA method for optimizing $\mathbf{v}$.

\subsection{Updating $\theta_{n}$}
The RIS phase shifts are the solutions of the problem:
\begin{equation}
\mathop {\max }\limits_{\left\{ {{\theta _n}} \right\}} - {\sum\limits_{n = 1}^N {\left| {{v_n} - {f_n}({\theta _n}){e^{j{\theta _n}}}} \right|} ^2}\;,\;\mathtt{s.t.} - \pi  \leqslant {\theta _n} \leqslant \pi \label{theta_problem1}
\end{equation}

It can be seen that the problem is separable over $n$, i.e., each summand can be optimized separately. Thus, defining $ \varphi_n=\arg ({v_n}) $, the optimal $\theta_{n}$ is found by solving the problem
\begin{equation}
  \begin{aligned}
 \mathop {\max }\limits_{{\theta _n}\in[-\pi,\pi]} ~&2{f_n}({\theta _n})\left| {{v_n}} \right|\cos ({\varphi _n} - {\theta _n}) - f_n^2({\theta _n})
 \end{aligned}\label{theta_problem}
 \end{equation}
which can be solved by standard numerical methods.

\subsection{Convergence and Complexity}
Finally, the overall algorithm to solve the optimization problem is obtained by iteratively optimizing the different optimization variables. Each iteration monotonically increases the objective value, which guarantees convergence. Moreover, the computational complexity is polynomial in the number of variables, since only the solution of convex problems is required\footnote{We recall that a convex problem can be solved with a complexity $\mathcal{C}=\mathcal{O}(L^{\eta})$, where $L$ is the number of variables and $1 \leq \eta\leq 4$ \cite{TalBook}.}. Thus, the complexity of optimizing $\{\rho_{k}\}$ is $\mathcal{O}(K^{\eta_{k}})$,  while the complexity of optimizing $\{\mathbf{w}_{k}\}$ and $\mathbf{v}$ are $\mathcal{O}(I_{w}(MK)^{\eta_{w}})$, and $\mathcal{O}(I_{v}N^{\eta_{v}})$, respectively, with $I_{w}$ and $I_{v}$ being the number of iterations of the SCA methods used to optimize $\mathbf{w}$ and $\mathbf{v}$. On the other hand, the optimal $\{\alpha_{I,k},\alpha _{E,k},\beta_{Ik},\beta_{Ek}\}$ are available in closed-form in \eqref{alpha_beta_I}, \eqref{alpha_beta_E} and thus the complexity required for their computation can be neglected. Similarly, the complexity of Problem \eqref{theta_problem1} can also be neglected, as it is linear in $N$. In fact, the problem can be decoupled over the $N$ optimization variables, and, for each $N$, the optimal $\theta_{n}$ is obtain by solving \eqref{theta_problem}. Thus, the overall complexity of the proposed method is given by $\mathcal{C}=I(\mathcal{O}(K^{\eta_{k}})+\mathcal{O}(I_{w}(MK)^{\eta_{w}})+\mathcal{O}(I_{v}N^{\eta_{v}}))$, where $I$ is the number of alternating optimization iterations to be run until convergence. The exponents of the polynomial are not available in closed-form, but it is known that they are upper-bounded by $4$ \cite{TalBook}. A typical value is $3.5$, which comes up when interior-point methods are used \cite{993483}.

\section{Numerical Results}\label{sec:Numerical Results}
For our numerical study, we consider an RIS-assisted  MISO communication system, in which the $M=8$ transmit antennas are arranged in a uniform linear array, and a set of $K=4$ UEs are considered. The UEs are  randomly and uniformly distributed within a disk of $1~m$ radius centered at $(5~m,5~m)$. The $N$-elements RIS is located at $(0~m,5~m)$. All channels are modeled as $X=L_x\left(\sqrt{\frac{\epsilon}{1+\epsilon}}\bar {X}^{LOS}+\sqrt{\frac{1}{1+\epsilon}}\bar {X}^{NLOS}\right)$, where $\bar{X}^{LOS}$ and $\bar{X}^{NLOS}$ are the line-of-sight (LOS) and non-LOS (NLOS) components, and $X$ is either $\mathbf{G}$, $\mathbf{h}_{r,k}$, or $\mathbf{h}_{d,k}$. The NLOS component follows the Rayleigh fading model, while the LOS component is $\bar{X}^{LOS}=\mathbf{a}_N(\theta^{AoA}) \mathbf{a}_{M}^H(\theta^{AoD})$, with:
  \begin{align}
  &\mathbf{a}_N(\theta^{AoA})=[1,e^{j\frac{2\pi d}{\lambda}\sin(\theta^{AoA})}~,~...~,e^{j\frac{2\pi d}{\lambda}(N-1)\sin(\theta^{AoA})}]^T\notag\\
  &\mathbf{a}_M(\theta^{AoD})=[1,e^{j\frac{2\pi d}{\lambda}\sin(\theta^{AoD})}~,~...~,e^{j\frac{2\pi d}{\lambda}(M-1)\sin(\theta^{AoD})}]^T\notag
  \end{align}
where $d$ and $\lambda$ are the inter-antenna separation and the wavelength, respectively. We assume $d/\lambda=1/2$. The large-scale distance-dependent path-loss is
$L=C_0{\left(\frac{d}{D_0}\right)^{-x_l}}$, where $d$ is the link distance and $D_0=1$ is the reference distance at which the reference path-loss $C_0=-30$dB is defined, $x_l$ is the path-loss exponent. The rest of the simulation parameters are given in Table 1. For comparison, we evaluate the performance gain achieved by the proposed algorithm in comparison with the ``No-RIS'' scenario, in which there is no RIS deployed in the system.
\begin{table}[!t]
\small
  \begin{center}
    \caption{Simulation Parameters.}
     \label{tab:table1}
       \begin{tabular}{|c|c|}
 \hline
\textbf{Parameters}  & \textbf{Values} \\
 \hline
  Number of RIS elements  & 60 \\
  \hline
  Maximum transmission power & $P_T$= 10 dB \\
  \hline
  Path-loss exponent - RIS-aided channels  & $x_{l}=2.2$ \\
  \hline
   Path-loss exponent - Direct channel & $x_{l}=3.6$ \\
  \hline
  Rician factor  & $\epsilon= 5$ dB \\
  \hline
 Power conversion noise & $\delta_{k}^2=\delta^2=- 50$dBm \\
  \hline
  Thermal noise power & $\sigma_{k}^2=\sigma^2=- 40$dBm  \\
  \hline
 Minimum harvested power & $P_{min}=10^{-5}$mW \\
  \hline
 Minimum SINR requirement & $\gamma_{\min}=10$dB \\
  \hline
  Power conversion efficiency  & $\eta_k=\eta= 0.6$ \\
  \hline
  Combining weight & $\bar\lambda$= [0.1, 1] \\
  \hline
  Conversion efficiency (uplink) $\xi_k$ & 0.005 \\
  \hline
\end{tabular}
\end{center} \vspace{-0.5cm}
\end{table}

\begin{figure*}[!t]
    \centering
    \begin{minipage}{0.47\linewidth}
        \centering
        \includegraphics[scale=0.55]{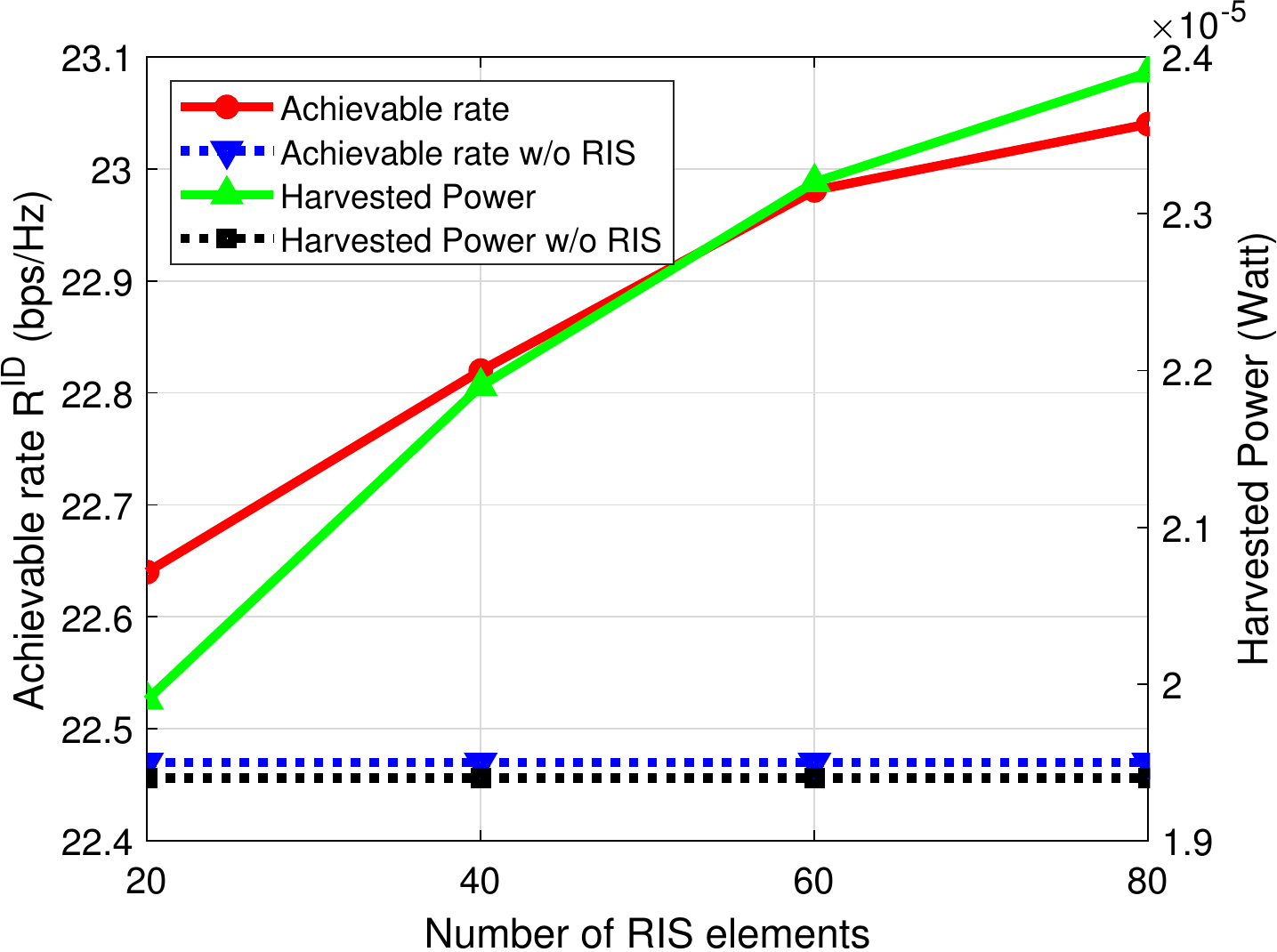}
        \caption{Sum-rate and harvested power versus the number of  RIS elements for $\bar \lambda=$0.6.} \label{R_HE_With/o RIS}
    \end{minipage} \vspace{-0.25cm} \hfill
    \begin{minipage}{0.47\linewidth}
        \centering
        \includegraphics[scale=0.55]{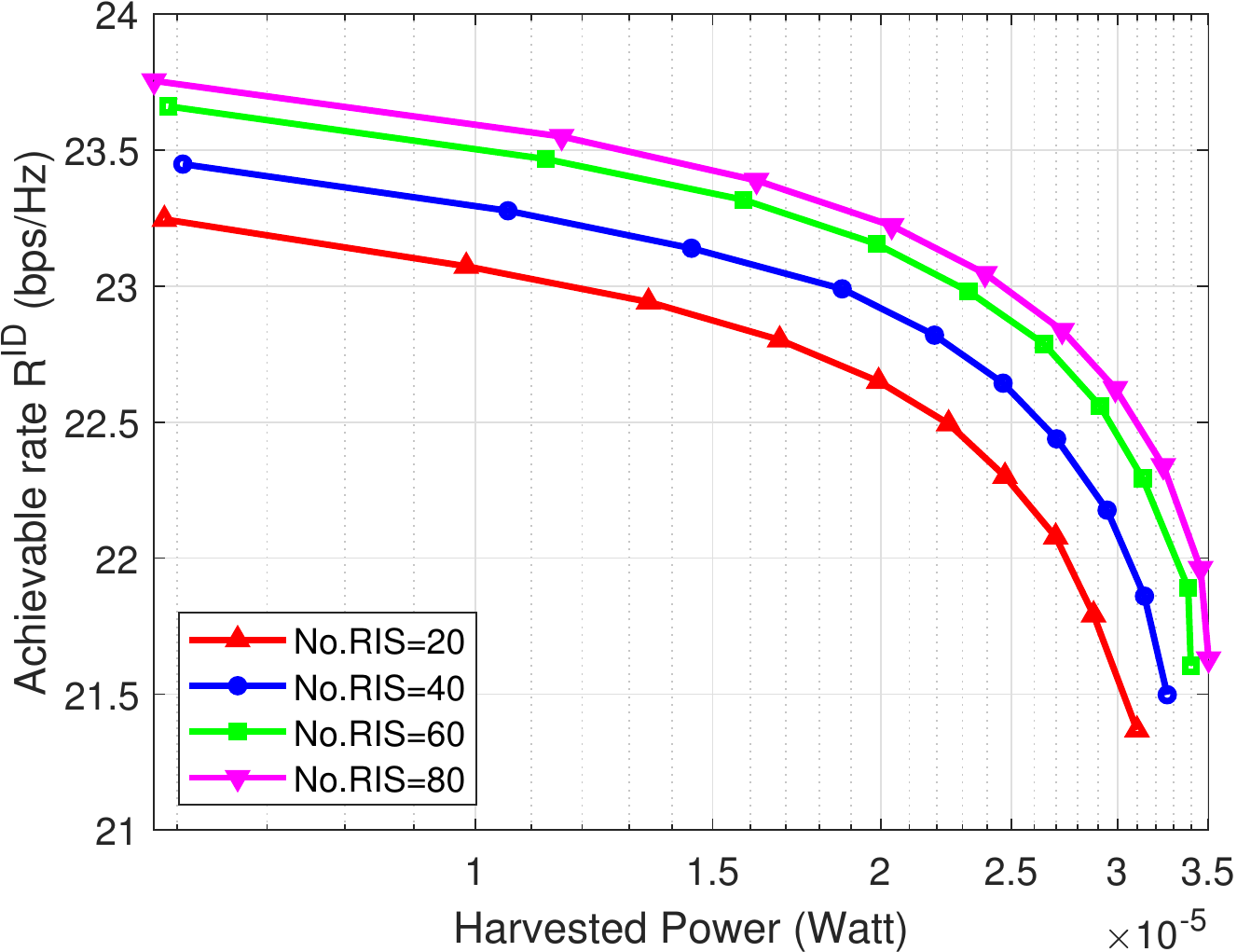}
        \caption{Trade-off between the sum-rate and harvested power with the number of RIS elements.} \label{NRIS_trade_off}
    \end{minipage} 
\end{figure*}

Figure \ref{R_HE_With/o RIS} depicts the relationship between the number of RIS elements versus the sum-rate and the harvested power. It can be observed that employing more RIS  elements leads to a monotonic growth of the amount of harvested power and sum-rate. The figure reveals the effectiveness of the proposed algorithm compared to the ``No RIS'' case (denoted by ``w/o RIS''). This monotonic gain is due to the appropriate design of the phase shift vector of the RIS  elements, which results in strong virtual LOS paths between the BS and the UEs.

In Fig. \ref{NRIS_trade_off}, we explore the trade-off between the sum-rate and harvested power as a function of the number of RIS elements. We observe that employing more RIS elements enhances the sum-rate and harvested power. Moreover, the figure highlights the impact of the preference parameter $\bar{\lambda}$. The preference parameter $\bar{\lambda}$ is utilized to determine the service priority between optimizing the sum-rate (i.e., with a low value of $\bar{\lambda}$) or optimizing the harvested power (i.e., with a high value of  $\bar{\lambda}$). When the system prioritizes power harvesting, the proposed algorithm allocates more power to the power harvesting receiver, and thus the sum-rate decreases. Similarly, reducing the value of $\bar{\lambda}$ gives higher priority to information decoding. The figures reports the trade-off between the sum-rate and harvested power for any values of $\bar{\lambda}$.

\begin{figure*}[!t]
    \centering
    \begin{minipage}{0.47\linewidth}
        \centering
        \includegraphics[scale=0.55]{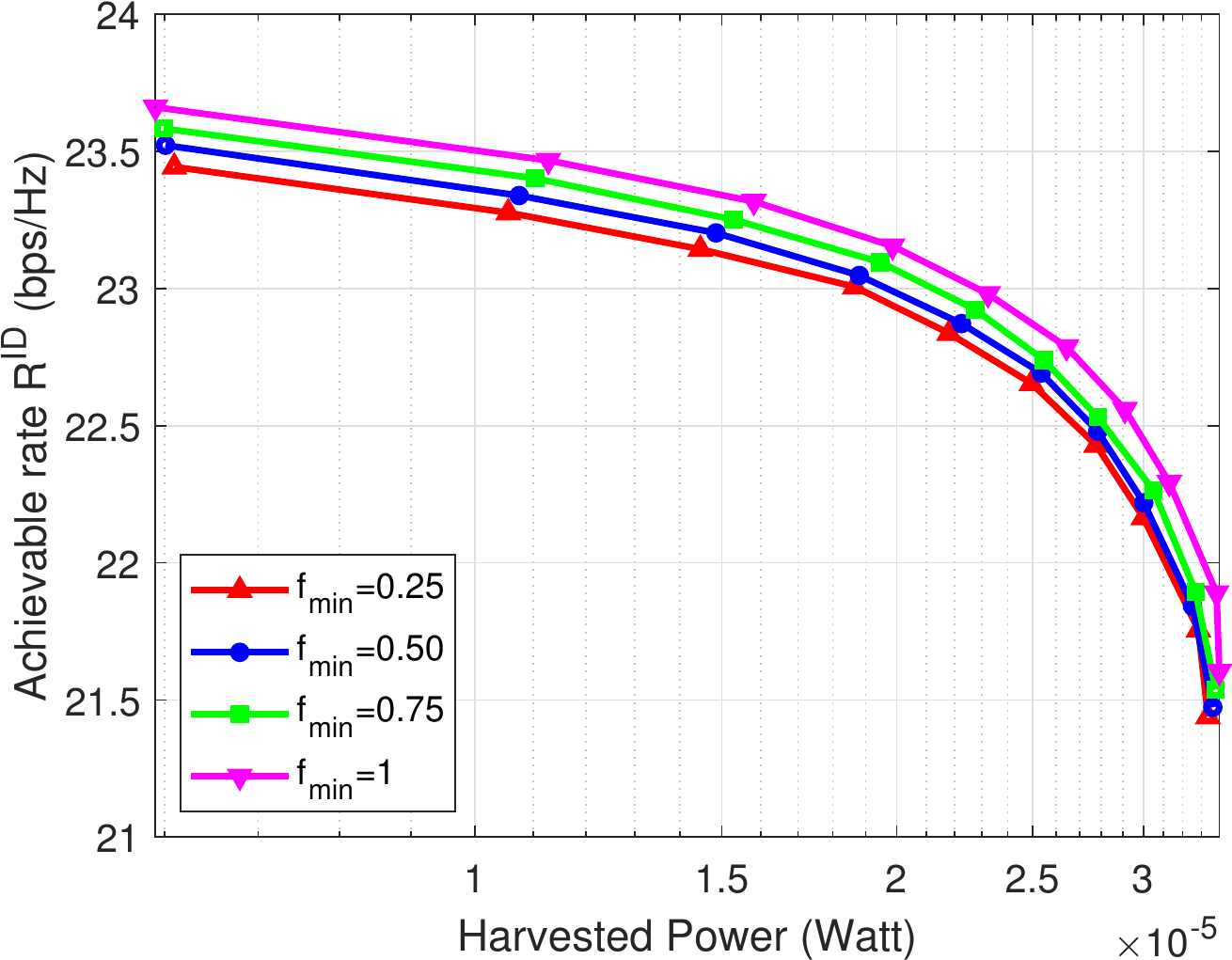}
        \caption{Trade-off between the sum-rate and harvested power with $f_{min}$.} \label{fmin_trade_off}
    \end{minipage} \vspace{-0.25cm} \hfill
    \begin{minipage}{0.47\linewidth}
        \centering
        \includegraphics[scale=0.55]{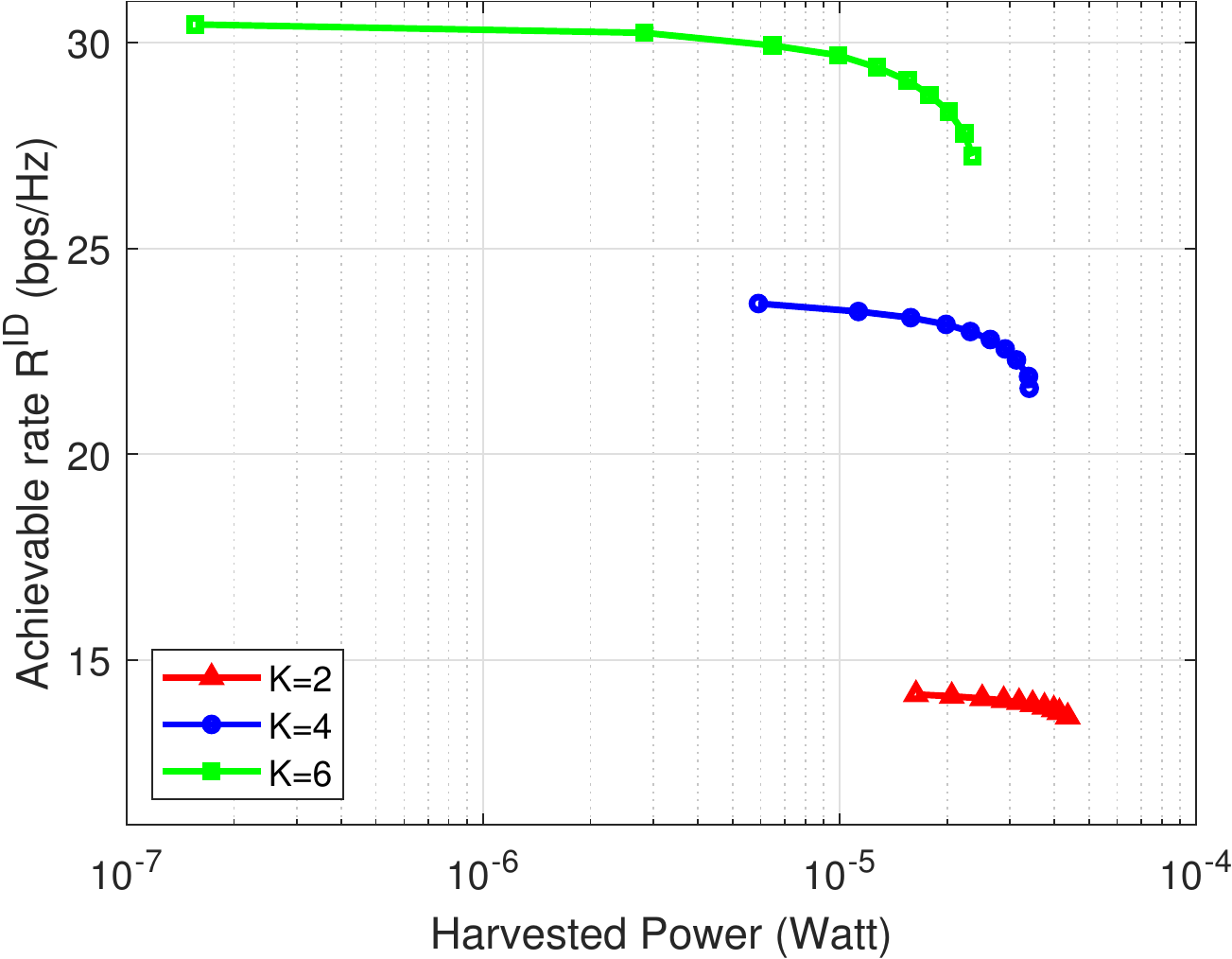}
        \caption{Trade-off between the sum-rate and harvested power with the number of UEs $K$.} \label{Nusers_trade_off}
    \end{minipage} 
\end{figure*}

In Fig. \ref{fmin_trade_off}, we examine the trade-off between the sum-rate and harvested power with the minimum value of the amplitude of the reflection coefficient $f_{min}$ of the RIS elements. We observe that the sum-rate and the harvested power increase as $f_{min}$ increases. The highest values are obtained when $f_{min}=1$, which is the ideal case for an RIS. In Fig. \ref{Nusers_trade_off}, finally, we investigate the trade-off between the sum-rate and harvested power as a function of the number of UEs $K$. We see that the sum-rate and harvested power increases and decreases, respectively, as the number of UEs increases. This can be explained because the sum-rate is proportional to $min(M, KN_r)$, where $N_r=1$ is the number of receive antennas and $M=8$ is fixed. Thus, the achievable data rate boosts as the number of UEs grows. On the other hand, the harvested power decreases due to the inverse relationship between the harvested power and the sum-rate.

\section{Conclusion} \label{sec:conclusion}
In this paper, we have investigated the trade-off between the sum-rate and power harvested in a multi-user RIS-aided downlink MISO system with SWIPT. Specifically, enforcing QoS constraints and practical phase shifts constraints, the transmit beamforming vector, the power splitting ratio, and the RIS reflection coefficients are jointly optimized by a two-layer penalty-based algorithm. Simulation results show that the proposed algorithm can significantly outperform conventional system deployments in the absence of RISs.

\bibliographystyle{IEEEtran}
\bibliography{libforcitation_NEW3}

\begin{thebibliography}{10}
\providecommand{\url}[1]{#1}
\csname url@samestyle\endcsname
\providecommand{\newblock}{\relax}
\providecommand{\bibinfo}[2]{#2}
\providecommand{\BIBentrySTDinterwordspacing}{\spaceskip=0pt\relax}
\providecommand{\BIBentryALTinterwordstretchfactor}{4}
\providecommand{\BIBentryALTinterwordspacing}{\spaceskip=\fontdimen2\font plus
\BIBentryALTinterwordstretchfactor\fontdimen3\font minus
  \fontdimen4\font\relax}
\providecommand{\BIBforeignlanguage}[2]{{%
\expandafter\ifx\csname l@#1\endcsname\relax
\typeout{** WARNING: IEEEtran.bst: No hyphenation pattern has been}%
\typeout{** loaded for the language `#1'. Using the pattern for}%
\typeout{** the default language instead.}%
\else
\language=\csname l@#1\endcsname
\fi
#2}}
\providecommand{\BIBdecl}{\relax}
\BIBdecl

\bibitem{RIS_GE_JSAC}
M.~Di~Renzo, A.~Zappone, M.~Debbah, M.-S. Alouini, C.~Yuen, J.~de~Rosny, and
  S.~Tretyakov, ``Smart radio environments empowered by reconfigurable
  intelligent surfaces: How it works, state of research, and the road ahead,''
  \emph{IEEE J. Sel. Areas Commun.}, vol.~38, no.~11, pp. 2450--2525, 2020.

\bibitem{Wu2019f}
Q.~Wu and R.~Zhang, ``Towards smart and reconfigurable environment: Intelligent
  reflecting surface aided wireless network,'' \emph{IEEE Commun. Mag.},
  vol.~58, no.~1, pp. 106--112, 2020.

\bibitem{huang2019holographic}
C.~Huang, S.~Hu, G.~C. Alexandropoulos, A.~Zappone, C.~Yuen, R.~Zhang, M.~D.
  Renzo, and M.~Debbah, ``Holographic {MIMO} surfaces for 6g wireless networks:
  Opportunities, challenges, and trends,'' \emph{IEEE Wirel. Commun.}, vol.~27,
  no.~5, pp. 118--125, 2020.

\bibitem{Wu2019k}
Q.~Wu and R.~Zhang, ``Joint active and passive beamforming optimization for
  intelligent reflecting surface assisted {SWIPT} under {QoS} constraints,''
  \emph{IEEE J. Sel. Areas Commun.}, vol.~38, no.~8, pp. 1735--1748, 2020.

\bibitem{Z.LiNOMASWIPT}
Z.~Li, W.~Chen, Q.~Wu, K.~Wang, and J.~Li, ``Joint beamforming design and power
  splitting optimization in {IRS}-assisted {SWIPT} {NOMA} networks,''
  \emph{IEEE Trans. Wireless Commun.}, vol.~21, no.~3, pp. 2019--2033, 2022.

\bibitem{Pan2019b}
C.~Pan, H.~Ren, K.~Wang, M.~Elkashlan, A.~Nallanathan, J.~Wang, and L.~Hanzo,
  ``Intelligent reflecting surface aided {MIMO} broadcasting for simultaneous
  wireless information and power transfer,'' \emph{IEEE J. Sel. Areas Commun.},
  vol.~38, no.~8, pp. 1719--1734, 2020.

\bibitem{xu2021optimal}
D.~Xu, V.~Jamali, X.~Yu, D.~W.~K. Ng, and R.~Schober, ``Optimal resource
  allocation design for large {IRS}-assisted {SWIPT} systems: A scalable
  optimization framework,'' \emph{IEEE Trans. Commun}, vol.~70, no.~2, pp.
  1423--1441, 2022.

\bibitem{khalili2021multiobjective}
A.~Khalili, S.~Zargari, Q.~Wu, D.~W.~K. Ng, and R.~Zhang, ``Multi-objective
  resource allocation for {IRS}-aided {SWIPT},'' \emph{IEEE Commun. Lett.},
  vol.~10, no.~6, pp. 1324--1328, 2021.

\bibitem{Chen_He_A_Joint_Power_Splitting}
C.~He, X.~Xie, K.~Yang, and Z.~J. Wang, ``A joint power splitting, active and
  passive beamforming optimization framework for {IRS} assisted {MIMO SWIPT}
  system,'' \emph{arXiv preprint arXiv:2105.14545}, 2021.

\bibitem{Abeywickrama2020a}
S.~Abeywickrama, R.~Zhang, Q.~Wu, and C.~Yuen, ``Intelligent reflecting
  surface: Practical phase shift model and beamforming optimization,''
  \emph{IEEE Trans. Commun}, vol.~68, no.~9, pp. 5849--5863, 2020.

\bibitem{Tang2020}
J.~Tang, Y.~Yu, M.~Liu, D.~K.~C. So, X.~Zhang, Z.~Li, and K.-K. Wong, ``Joint
  power allocation and splitting control for {SWIPT}-enabled {NOMA} systems,''
  \emph{IEEE Trans. Wireless Commun.}, vol.~19, no.~1, pp. 120--133, 2020.

\bibitem{Shen2018}
K.~Shen and W.~Yu, ``Fractional programming for communication systems—part
  ii: Uplink scheduling via matching,'' \emph{IEEE Trans. Signal Process.},
  vol.~66, no.~10, pp. 2631--2644, 2018.

\bibitem{993483}
S.~Boyd and L.~Vandenberghe, \emph{Convex optimization}.\hskip 1em plus 0.5em
  minus 0.4em\relax Cambridge university press, 2004.

\bibitem{8579566}
C.~Pan, H.~Ren, M.~Elkashlan, A.~Nallanathan, and L.~Hanzo, ``Robust
  beamforming design for ultra-dense user-centric {C-RAN} in the face of
  realistic pilot contamination and limited feedback,'' \emph{IEEE Trans.
  Wireless Commun.}, vol.~18, no.~2, pp. 780--795, 2019.

\bibitem{TalBook}
A.~Ben-Tal and A.~Nemirovski, \emph{Lectures on Modern Convex Optimization:
  Analysis, Algorithms, Engineering Applications}.\hskip 1em plus 0.5em minus
  0.4em\relax MPS-SIAM Series on Optimization, SIAM, 2001.

\end{thebibliography}

\end{document}